\documentclass[twocolumn,twoside,slac_two]{revtex4_MOD}
\usepackage{graphicx}
\usepackage{fancyhdr}
\def\fermi{\textsl{Fermi}}
\def\fermilat{\textsl{Fermi}/LAT}
\def\agile{\textsl{AGILE}}
\def\swift{\textsl{Swift}}

\begin{document}

\title{High-resolution monitoring of parsec-scale jets in the \textsl{Fermi} era}

%

\author{Eduardo Ros}
\affiliation{Departament d'Astronomia i Astrof\'{\i}sica, Universitat de Val\`encia, E-46100 Burjassot, Valencia, Spain\\ Max-Planck-Institut f\"ur Radioastronomie, Auf dem H\"ugel 69, D-53121 Bonn, Germany}

\begin{abstract}
I review here the present observational efforts to 
study parsec-scale radio jets in active galactic nuclei with
very-long-baseline interferometry (VLBI) as related to the 
new window to the Universe opened by the LAT instrument on-board
the \fermi\ \textsl{Gamma-Ray Space Telescope}.
I describe the goals and achievements of those radio studies, 
which aim to probe the emission properties, morphological changes 
and related kinematics, magnetic fields from the linear and circular 
polarization, etc., and I put those in the context of the 
radio--gamma-ray connection.  Both statistical studies based on
radio surveys and individual studies on selected sources are reported.
Those should shed some light in the open questions about the nature of
emission in blazars.
\end{abstract}

\maketitle

\thispagestyle{fancy}


\section{INTRODUCTION}
Since the measurements from \textsl{CGRO}/EGRET we know that the 
$\gamma$-sky is dominated by 
the Galactic plane diffuse emission, pulsars, and blazars.  
This has been confirmed
with the outstanding findings of \fermi\ and its LAT detector,
launched in June 2008 and operative since August 2008. 
The 2nd \fermilat\ catalog (2LAC) 
\cite{ack11a} includes all extragalactic
sources with a significant detection
over the first two years of scientific operation. The so-called `clean sample' of the 2LAC contains 
886 sources, from which 
310 are flat-spectrum radio quasars (FSRQ), 395 are BL\,Lac objects, 
157 candidate blazars of unknown type, 8 misaligned active galactic
nuclei (AGN), 4 narrow-line Seyfert\,1 (NLS1), 10 AGN of other types, and 2 starburst galaxies.  
Notice that sources with only sporadic activity were missing,
since they don't reach the test statistic threshold of 25 
($\mathrm{TS}>25$)\footnote{Remarkably, sources like 3C\,120 or 3C\,111, which
were present in earlier catalogs, are not listed at the 2LAC.}.


\fermilat\ has shown that BL\,Lacs are the most common $\gamma$-emitters, 
more frequently than FSRQ.  One should be aware of the
strong biases introduced by Doppler boosting on the observed flux
from AGN, which biases the brightest extra-galactic $\gamma$-ray objects and
high frequency radio sources towards fast jets with a small viewing
angle, that is, close to the line of sight (e.g., \cite{lis10a,lis10b}).
The picture is completed by observations of the \agile\ 
$\gamma$-mission \cite{tav09} (2008-)
the rapid AGN results provided by the X-ray and $\gamma$-ray mission 
\swift\ \cite{geh04} (2005-)
and the ground-based very-high-energy (VHE) 
$\gamma$-ray Cherenkov telescopes such as 
HESS (2003-),
MAGIC (2004-),
CANGAROO-III (2004-),
or
VERITAS (2006-).

Among the big questions raised in earlier editions of this workshop,
it is open if $\gamma$-ray flares originate in relativistic
shocks, what is the distance of the main energy dissipation site from the
central engine, what are the emission mechanisms at stake, and what
relates the brightness in the radio with the $\gamma$-rays.  
The observational tools to address the \fermi\ era are 
VLBI campaigns to be related with the continuous all-sky $\gamma$-ray 
observations
by \fermilat, complemented
with multi-band campaigns including as well IR, optical, UV, and X-rays.
The information achieved by intensive flux density monitoring 
campaigns are being addressed by other authors at this conference.

To review this topic, I will first introduce the observables measured
directly and indirectly by VLBI and the other parameters to be 
compared with from the $\gamma$-ray monitoring and from the spectral
energy distribution (SED) studies of AGN. 
I will continue describing the main survey campaigns
and some of their highlights so far.  To complement this,
I will present a (necessarily incomplete) selection of studies 
on individual sources combining VLBI and $\gamma$-observations.

\section{WHAT IS MEASURED BY VLBI?}

\subsection{VLBI targets} 
Blazars display powerful jets oriented towards the observer, and show
high brightness temperatures in the radio regime, which allow them to
be observed by VLBI.  
VLBI is a technique that provides resolutions of the order of the
milliarcsecond (parsecs at cosmological distances), working regularly
from 3\,mm up to 1\,m wavelengths.  At the longest wavelengths the
presence of the ionosphere as a dispersive propagation medium distorts
the waves and limits VLBI performance.  At the shortest wavelengths, where
the highest resolution is achieved, 
atmospheric turbulence, and especially water vapour, disturb the
observations, causing coherence loss and limiting the integration time
and therefore its detection threshold.  
Wavelengths from 7\,mm up to 18\,cm are the most commonly used.  

One of the main targets of VLBI are AGN, given their high 
brightness temperatures $T_\mathrm{b}$, of up to $10^{12}$\,K in 
the core, dropping to $10^{10}$\,K or lower values in the jet 
(see below for the definition of $T_\mathrm{b}$). 

Distinct features or `blobs' in the jet can be identified with shocks or
instabilities in the jet.  The magnetic field orientation can be estimated
by the linear and circular polarisation, if observing in this mode.  
The structural changes observed by combining several epochs can
be associated with helical jets (e.g., \cite{sav06}) or binary black 
holes (e.g., \cite{lob05}).  Identifying 
`moving' features from different observing times, the kinematics of
those components are established, and even the ejection times of
the features at the base of the jet could be related to the 
outburst observed at the single-dish light curves of the 
sources \cite{sav02}.
At the base of the jet, the emission is self-absorbed by synchrotron,
so that the peak of brightness (also labeled as jet 'core') corresponds
to different physical locations at different frequencies.  The
absolute position of the source can be recovered by
astrometric methods \cite{ros05}.  If this is not possible,
images at different frequencies can be registered by aligning the optically
thin regions either by model fitting with Gaussian functions \cite{kad04},
by cross-correlation of the jet features \cite{osu09,fro09}, 
or by a combination of astrometry and jet alignment \cite{gui95,ros01}.
After the core-shift correction, the synchrotron turnover frequency 
and flux density can be computed
all over the jet, and from those, physical parameters in the jet are estimated, such as the
magnetic field, and pressure gradients \cite{lob98}.

\subsection{Extracting information from VLBI images}
So, from one single VLBI image, or combining them in time (kinematics)
or frequency (spectral studies), we can measure directly several 
physical quantities in AGN,
most of them affected the observed flux 
by Doppler beaming, caused by relativistic
effects and the small viewing angle of the jet, pointed almost
towards the observer.
Given a region of the jet moving downstream with a speed $\beta=v/c$ 
in the rest frame ($c$ is
the light speed), with an angle $\theta$ between the jet,
the Lorentz factor will be
$\Gamma=(1-\beta^2)^{-1/2}$
and
the Doppler factor will be
$\delta = (\Gamma(1-\beta\cos\theta))^{-1}$.
Beaming will affect several magnitudes, and the relativistic effects
and the small viewing angle will affect several parameters being
measured by VLBI.  I summarize those in Table~\ref{table:VLBI+gamma},
and describe them in the following paragraphs.

\begin{table}[t]
\begin{center}
\caption{Summary of observable VLBI, 
SED, and $\gamma$-ray parameters 
\label{table:VLBI+gamma}}
\begin{tabular}{|l|c|c|}
\hline 
\multicolumn{3}{|c|}{\textbf{Radio}} \\
\hline 
\multicolumn{2}{|l|}{Parameter} & Units \\
\hline
Radio detection		&  & \\ 
Apparent speed 		& $\beta_\mathrm{app}$  & $c$ \\
Flux density		& $S$			& Jy \\
Brightness temperature 	& $T_\mathrm{b}$	& K \\
Apparent opening angle	& $\psi$		& deg \\
Luminosity		& $L_\mathrm{R}$	& W\,Hz$^{-1}$ \\
Jet-to-counterj. ratio	& R			& -- \\
P.A.\ misalignment$^\mathrm{a}$	
			& $\Delta\phi$		& deg \\
Spectral index		& $\alpha$		& -- \\
Polarisation angle	& $\chi$		& deg \\
Polarisation level	& $m$			& \% \\
Faraday rotation	& RM			& rad\,m$^{-2}$ \\
\hline
Viewing angle		& $\theta$		& deg \\
Lorentz factor		& $\Gamma$		& -- \\
Doppler factor		& $\delta$		& -- \\
Ejection epoch		& $t_0$			& yr \\
\hline
\multicolumn{3}{|c|}{\textbf{SED}} \\
\hline
Sync.\ frequency peak 	& $\nu_\mathrm{max,sync}$ & Hz \\
Lum.\ at sync.\ peak 	& $L_(\nu=\nu_\mathrm{max, sync})$
						& W\,Hz$^{-1}$  \\
Inv.\ Compton frequency peak
			& $\nu_\mathrm{max,IC}$	& Hz \\
Lum.\ at inv.\ Compton peak
			& $L_(\nu=\nu_\mathrm{max,IC}$
						& W\,Hz$^{-1}$  \\
\hline
\multicolumn{3}{|c|}{\textbf{Gamma}} \\
\hline
$\gamma$-detection	&  			& \\ 
Flare epoch		& $t_{\gamma\mathrm{-flare}}$
						& yr \\
Flux			& $S_\gamma$		& Jy \\
Flux variability	& $\Delta S_\gamma/S_\gamma$		
						& -- \\
Luminosity		& $L_\gamma$		& W\,Hz$^{-1}$ \\
Photon index		& $\Gamma_\gamma$	& -- \\
\hline
Gamma-ray to radio flux ratio	& $G_r$			& -- \\
\hline 
\multicolumn{3}{l}{$^\mathrm{a}$ Kiloparsec- and parsec-scale misalignment}\\
\end{tabular}
\end{center}
\end{table}

\subsubsection{Directly measured parameters by VLBI}
After identifying features between different observing epochs, we can
determine the sky motion of those and compute the 
\textbf{apparent speed} 
$\beta_\mathrm{app}=\beta\sin\theta/(1-\beta\cos\theta)$.
For a given $\beta$, the maximum speed 
$\beta_\mathrm{app,max}=\beta\Gamma$  
is reached when 
$\cos\theta_\mathrm{max}=\beta$.

After hybrid mapping or Gaussian model fitting, we can measure
for each feature the 
value of the flux density $S$, usually expressed in Jy 
($10^{-26}$\,W\,m$^{-2}$\,Hz$^{-1}$). The \textbf{luminosity}
can be obtained directly from the flux density:
$L = 4\pi D_L^2 S$
where $D_L$ is the luminosity distance (to be computed from the redshift $z$
measured in the optical).  The intrinsic and observed luminosity
are also affected by the $K$-correction and 
Doppler boosting (and the spectral index
$\alpha$, from $S\propto \nu^{+\alpha}$) as follows
$L_\mathrm{obs}=L_\mathrm{int}\times\delta^{n-\alpha}\times (1+z)^{-(1-\alpha)}$ 
($n=2\,,3$).

Having the flux density $S$ and the interferometer 
resolution (beam), we compute the
\textbf{brightness temperature} 
$T_\mathrm{b}=1.222\times10^{12} S(1+z) / \nu^2 a b$
where $T_\mathrm{b}$ is given in K, 
$\nu$ is the observing frequency in Hz, $S$ is the flux density in 
Jy, 
and $a$ and $b$ are the major and minor beam axes, respectively, in 
milliarseconds.
The intrinsic and observed brightness temperatures 
are related with the beaming Doppler factor $\delta$ by
$T_\mathrm{b,obs} = T_\mathrm{b,int} \times \delta$.

We can also measure the difference in position angle of the
jet in parsec- and kiloparsec-scales (the latter from connected
interferometers such as the VLA or MERLIN), and get the
\textbf{jet misalignment angle} $\Delta\phi$. 

For images with a high dynamic range
we can obtain the \textbf{jet-to-counterjet} ratio, which 
is $R=((1+\beta\cos\theta)/(1-\beta\cos\theta))^{2-\alpha}=(\beta_\mathrm{app}^2+\delta^2)^{2-\alpha}$.

By measuring how the jet gets
broader (see e.g., \cite{pus09,ojh10}), we measure the 
\textbf{apparent opening angle} $\psi$.
Since the jets are not precisely in  the plane of the sky, 
the apparent and the intrinsic 
opening angle are related by
$\psi_\mathrm{int} = \psi_\mathrm{obs}\sin\theta$.

Finally, if we observe in polarisation mode, we can measure
the \textbf{polarisation level} $m$ (as the ratio 
of linearly polarised
and total intensity)  as well as the \textbf{polarisation 
angle} $\chi$ (also known as electric vector position angle or EVPA).
If we have several frequencies, we can compute the change of the
EVPA as a function of the squared wavelength and determine the
Faraday \textbf{rotation measurement} RM.


\subsubsection{Indirectly measured parameters}
From the parameters measured above we can 
see that the intrinsic parameters $\beta$, $\theta$, or $\delta$
are degenerate and several solutions are possible for 
a given value of $R$, $\beta_\mathrm{app}$, etc.  

We can measure $\delta$ independently of the VLBI kinematical
analysis by computing the flux density variations measured
by a densely sampled single-dish or VLBI monitoring campaign \cite{hov09}:
The variability time can be defined from the variations of the
flux density $S$ 
by decomposing a flux density flare into exponential flares as 
$\Delta S(t) = \Delta S_\mathrm{max} e^{(t-t_\mathrm{max})/\epsilon \tau}$,
where $\Delta S_\mathrm{max}$ is the maximum amplitude of the flare in Jy,
$t_\mathrm{max}$ is the epoch of the flare maximum and $\tau$ is the
rise time of the flare.  $\epsilon=1$ for $t<t_\mathrm{max}$ and
$\epsilon=1.3$ for $t>t_\mathrm{max}$.  In this way,
$T_\mathrm{b,obs(var)} = 1.474\times 10^{13} S_\mathrm{max} D_L^2 \nu^{-2} 
\tau^{-2} (1+z)^{-1}$
where $D_L$ is the luminosity distance in Mpc and $\nu$ the frequency in
Hz.
From here
the \textbf{variability Doppler factor} is
$\delta_\mathrm{var} = (T_\mathrm{b,obs(var)}/T_\mathrm{b,int})^{1/3}$,
where $T_\mathrm{b,int}$ is assumed to be $5\times10^{10}$\,K.

With the apparent speed $\beta_\mathrm{app}$ 
and  $\delta_\mathrm{var}$, we also obtain the bulk Lorentz factor
$\Gamma=(\beta_\mathrm{app}^2 + \delta_\mathrm{var}^2 +1)/(2\delta_\mathrm{var})$ and the 
\textbf{viewing angle}
$\theta=\tan(2\beta_\mathrm{app}/\beta_\mathrm{app}^2 + \delta_\mathrm{var}^2 -1)$.  Alternatively, an upper bound for the viewing angle can be obtained
from the jet-to-counterjet ratio as 
$\theta<\beta^{-1}\arccos(R^{1/2-\alpha}-1/R^{1/2-\alpha}+1)$.

Last but not least,
knowing the apparent speed $\beta_\mathrm{app}$
we can compute the \textbf{ejection time} $t_0$ simply by extrapolating the time
for which the distance of the feature to the core is zero.  
If jet features are related to plasma injections at the base of the
jet, this should be seen at high frequencies, as it has
been observed in X-rays for 3C\,120 \cite{mar02} or NGC\,1052 \cite{ros08}.

\subsection{SED properties}
AGN SED (representation of intensity as a function of frequency) show usually two bumps, caused by 
non-thermal synchrotron emission at the low energies and (most probably)
by inverse Compton up-scattering of ambient optical-UV photons, although the contribution from energetic hadrons cannot be ruled out (e.g., \cite{urr11}).
Therefore, the \textbf{positions and luminosities of
SED peaks} of both bumps are also used for correlation studies:
$(\nu_\mathrm{max,sync},L_(\nu=\nu_\mathrm{max, sync})$ and 
$(\nu_\mathrm{max,IC},L_(\nu=\nu_\mathrm{max,IC})$.  From this,
additional parameters such as the \textbf{gamma-radio loudness}
$G_r= L_\gamma / L_R$ has been defined \cite{lis11}.

\subsection{Correlating VLBI measurements and $\gamma$-properties}
Most of the statistical studies with radio surveys are based in 
relating the above mentioned parameters with the $\gamma$-measurements.
The first check to be performed are cross-correlating catalogs (radio with the 
\fermi\ ones--three-month list \cite{abd09b}, one-year 
AGN catalog \cite{abd10c}, or the two-year AGN catalog \cite{ack11a}), and
comparing $\gamma$-detection with radio properties.
After that, the values of the radio parameters 
can be plotted against the high-energy parameters, and correlations
are searched by using different statistical methods.  

I have listed the parameters from $\gamma$-ray observations in
Table~\ref{table:VLBI+gamma} as well.
In principle, those high-energy parameters are the 
\textbf{$\gamma$-flux} $S_\gamma$ and its \textbf{$\gamma$-flux variability}
$\delta S_\gamma/S_\gamma$, the 
\textbf{$\gamma$-luminosity}\footnote{Notice that computing 
the luminosity in the $\gamma$-regime is more
problematic than in the (almost monochromatic) radio regime, 
given the fact that photons with frequencies different in
several orders of magnitude are being used.  
An expression for the $\gamma$-luminosity is given in \cite{lis11} 
as $L_\gamma = 4\pi D_L^2S_{0,\,1} / (1+z)^{2-\Gamma_\gamma}$, 
where 
$S_{0,\,1}= C_1 E_1 F_{0,\,1} (\Gamma_\gamma-1/\Gamma_\gamma-2) 
(1-(E_1/E_2)^{\Gamma_\gamma-2})$,  
being $F_{0,\,1}$ the upper limit on
photon flux above a given energy $E_1=0.1$\,GeV, the upper energy 
$E_2=100$\,GeV, and $C_1=1.602\times10^{-3}$\,erg/GeV\,=1\,J/J as
conversion factor. The authors fixed $\Gamma_\gamma=2.1$.}
$L_\gamma$, and the
\textbf{$\gamma$ photon index} 
$\Gamma_\gamma$ (since the number of photons per unit time per unit
area in a frequency bandwidth is $dN/dE=(F_\nu/h\nu) \nu_0$, where
$\nu_0=h/E_0$, where $E_0=1$\,keV and $h$ is Planck's constant,
from the radio astronomy convention, $S=F_\nu \propto \nu^{+\alpha}$,
$dN/dE \propto \nu^{\Gamma_\gamma} \propto \nu^{\alpha -1}$.
So, $\Gamma_\gamma=\alpha-1$ if we want
to compare the radio spectral index and the high energy photon index.
We can also add the flaring activity, including
\textbf{flare epochs} $t_{\gamma\mathrm{-flare}}$.

A naive approach is to check the relationship between the different
observables both in VLBI and in $\gamma$-rays, and draw 
physical conclusions
of them.  Having a sample of objects at one band, the 
properties at the other band can be divided between detections and
non-detections in histograms, and usually the sources are then divided
into their optical classification (FSRQ, BL\,Lac, Radio Galaxy, etc.)
or into their high-energy peak classification (HSP, ISP, LSP).
This was especially the approach on the first VLBI-related publications
of the \fermi\ era.  When more
data have been available, plots of properties at one band versus the 
other band provide some hints of the nature of radio-loud/quiet
and $\gamma$-loud/quiet objects.  

\section{VLBI SURVEYS}
VLBI has been performed since the early 1970s, and more 
intensively since the construction
of the Very Long Baseline Array (VLBA\footnote{The VLBA is operated by the US National
Radio Astronomy Observatory, a facility of the US National Science Foundation 
operated under cooperative agreement by Associated Universities, Inc.})
in the early 1990s.
Regular observations with open calls are being performed
by the European VLBI
Network (EVN\footnote{The European VLBI Network is a joint facility of European, Chinese, South African 
and other radio astronomy institutes funded by their national research councils.})
and the Long Baseline Array 
(LBA\footnote{The Long Baseline Array is part of the Australia Telescope which is
funded by the Commonwealth of Australia for operation as a National
Facility managed by CSIRO.}; expanded with the 
addition of telescopes outside Australia) operate regularly.
Several big surveys have monitored the brightest AGN for 
decades, every one with a different approach (a review on
VLBI imaging surveys is presented in \cite{kov09}).
Reaching both hemispheres, the geodetic networks, at present under
the umbrella of the International VLBI Service, collected
data of hundreds of sources for calibration purposes, and
for determining tectonic motions, Earth Orientation Parameters,
the length of the day, and other geophysical parameters. 

\begin{table}[t]
\caption{VLBI surveys complementing $\gamma$-observations
\label{table:surveys}}
\[
\centering
\resizebox{0.99\columnwidth}{!}{%
\begin{tabular}{|l|c|c|c|c|l|}
\hline 
\textbf{Program}		
		& \textbf{$\lambda$}
			& \textbf{$N_\mathrm{sources}$}
				& \textbf{$N_\mathrm{epochs}$$^\mathrm{a}$ }
					& \textbf{Time}
						& \textbf{Ref.} \\
\hline
GMVA 3mm	& 3\,mm	& 121	& 2	& 2004-	& \cite{lee08} \\
\textbf{Boston Univ.} 	
		& 7\,mm	& 35	& 50	& 2007-	& \cite{mar11} \\
TeV Sample	& 7\,mm$^\mathrm{b}$
			& 7	& 5	& 2006- & \cite{pin10} \\
\textbf{MOJAVE/2\,cm Survey}		
		& 2\,cm	& 300	& 20	& 1994-	& \cite{lis09a} \\
Bologna low-$z$	& 2/3.6\,cm 
			& 42	& 2	& 2010-	& \cite{gir10} \\
\textbf{TANAMI} 		
		& 1.3/3.6\,cm
			& 80	& 5	& 2008-	& \cite{ojh10} \\
\textbf{VIPS} 		
		& 6\,cm	& 1127	& 1	& 2007	& \cite{hem07} \\
VIPS subsample	& 6\,cm	& 100	& 2	& 2010-	& \cite{tay10} \\
CJF 		& 6\,cm & 293	& 3	& 1990s & \cite{pol03} \\
ICRF		& 3.6/13\,cm & 500 & 10 & 1990s & \cite{ojh04} \\
VCS		& 3.6/13\,cm & 3400 & 1 & 1990s & \cite{kov07} \\
\hline 
\multicolumn{6}{l}{\footnotesize $^\mathrm{a}$ Typical number of epochs per source} \\
\multicolumn{6}{l}{\footnotesize $^\mathrm{b}$ Also including $\lambda$1.3\,cm \& $\lambda$3.6\,cm}\\
\end{tabular}
}
\]
\end{table}

With astronomical goals, the Caltech-Jodrell Bank Survey, 
initiated in the early 1990s, was the first big effort, and
later it was followed by the 2\,cm VLBA Survey---now
turned into the MOJAVE project---in the mid 1990s, and 
newer projects with additional purposes such as the Boston
University blazar monitoring program, the 86\,GHz
Survey with the Global Millimetre VLBI Array, joined the
list.  During the 2000s, several projects were designed
to collect information for the 
\fermi\ era.  
I will list in the next sections some of the findings of the
ongoing surveys (summarised in Table~\ref{table:surveys}) reported
in refereed journals.
A summary of the correlations found is presented  in 
Table~\ref{table:VLBI-gamma-corr}.

\begin{table}[t]
\caption{Correlations reported in survey studies 
following the notation of table~\ref{table:VLBI+gamma}
\label{table:VLBI-gamma-corr}}
\[
\centering
\resizebox{0.99\columnwidth}{!}{%
\begin{tabular}{|c|c|c|c|c|c|c|c|c|} 
\hline 
R/$\gamma$ 	&
		Det? 	& 
			$t_{\gamma\mathrm{-f}}$ & 
				$S_\gamma$	&  
					$L_\gamma$ 	& 
						$\Gamma_\gamma$	& 
							$\delta S_\gamma/S_\gamma$ &
								$G_r$ &
									$\nu_\mathrm{m,s}$ \\ 
\hline
Det? 		& \cite{lin11}$^\mathrm{a}$
			&	&	&	&	&	&	& \\	
$\beta_\mathrm{app}$  
		& \cite{lis09b},\textsl{\cite{kar11}}
			&	&	& \cite{kar11}	
						&	& \cite{lis09b}	
								&	& \\	
$S_\mathrm{R}$	& \cite{kov09b} 	
			& \cite{pus10}$^\mathrm{d}$
				& \cite{kov09b,pus10},\textsl{\cite{lin11}}
					&	&	&	&	& \cite{lis11}$^\mathrm{e}$ \\
$T_\mathrm{b}$	& \cite{kov09b,lin11},\textsl{\cite{ojh10}}
			&	& \textsl{\cite{lin11}}	
					&	&	&	&	& \cite{lis11}$^\mathrm{f}$ \\
$\psi$		& \cite{pus09,ojh10,lin11}$^\mathrm{b}$
			&	&	&	&	&	& \cite{lis11} 
									& \\ %
$\Delta\phi$	& \cite{lis09b} 	
			&	&	&	&	&	&	& \\ 
$m$		& \cite{lin11} 	
			&	&	&	&	&	&	& \cite{lis11}$^\mathrm{g}$	 \\
\hline
$\theta$	& \cite{sav10}$^\mathrm{c}$
			&	&	&	&	&	&	& \\ 
$\Gamma$	& \textsl{\cite{sav10}} 	
			&	&	&	&	&	&	& \\ 
$\delta$	& \cite{sav10,lin11}	
			&	&	&	&	&	&   	& \\ 
\hline
$G_r$		&	&	&	&	& \cite{lis11}$^\mathrm{e}$
							&	&	& \cite{lis11} \\   		
\hline 
\multicolumn{9}{l}{\footnotesize \textsc{Note:} No correlation is shown in \textsl{italics}.} \\[-2pt]
\multicolumn{9}{l}{\footnotesize $^\mathrm{a}$ No correlation for BL\,Lac, dependent on $\delta$ for FSRQ} \\[-2pt]
\multicolumn{9}{l}{\footnotesize $^\mathrm{b}$ No correlation is found in \cite{pus09} for the intrinsic opening angle $\psi_\mathrm{int}$} \\[-2pt]
\multicolumn{9}{l}{\footnotesize $^\mathrm{c}$ The $\theta_\mathrm{int}$ distribution is narrower for the $\gamma$-detected sources in \cite{sav10}} \\[-2pt]
\multicolumn{9}{l}{\footnotesize $^\mathrm{d}$ The radio core flux is delayed w.r.t.\ $\gamma$-flares, following \cite{pus10}} \\[-2pt]
\multicolumn{9}{l}{\footnotesize $^\mathrm{e}$ Negative correlation} \\[-2pt]
\multicolumn{9}{l}{\footnotesize $^\mathrm{f}$ HSP BL\,Lac objects tend to be less compact than other objects} \\[-2pt]
\multicolumn{9}{l}{\footnotesize $^\mathrm{g}$ HSP BL\,Lac objects tend to lower polarisation levels.} \\[-2pt]
\end{tabular}
}
\]
\end{table}

\subsection{VIPS}
The VLBA Imaging and Polarimetry Survey\footnote{\texttt{http://www.phys.unm.edu/$\sim$gbtaylor/VIPS/}} is a one-epoch survey including
polarimetry at 5\,GHz perfomed in the mid 2000s.  The first stage of
the observations was described in \cite{hem07}. Results on the radio properties
of sources
detected by \fermilat\ showed no correlation between $S$ and $S_\gamma$ 
\cite{lin11}.  Furthermore, from this study radio-bright
BL\,Lac objects detected by
\fermilat\ were similar to the non-LAT ones, but for FSRQ there is a difference
on the emission related to Doppler boosting: not surprisingly, only the 
FSRQ with higher $\delta$ are $\gamma$-loud.  Polarisation at the
base of the jet is a signature as well for $\gamma$-ray loud AGN.

\subsection{Boston University Blazar Program}
The Boston University (BU) blazar group has been performing a monitoring
campaign with VLBA images sampled monthly at 43\,GHz since 2007 (continuing
the monitoring performed on some sources on the early 2000s), whose
calibrated data are publicly 
available\footnote{\texttt{http://www.bu.edu/blazars/VLBAproject.html}}.  
Due to the nature of the survey (reduced number of sources,
intensively monitored), the published results are 
focused on multi-band studies of individual sources (see below).
A description of the overall project is given in 
\cite{mar11}.  In general it is interpreted that 
a high $\gamma$-state is related to an outburst
at the millimetre regime.  The outburst is 
associated to the passing of a traveling shock 
through a recollimation shock in the base of jet
\cite{jor11}.  Other explanations are possible, e.g., a
pinch instability in a helical jet.

\subsection{MOJAVE}
The 2\,cm Survey 
project \cite{zen02} was started in 1994, short after
VLBA completion, 
with the aim of monitoring a sample
of bright and representative AGN at sub-parsec scales.
It was continued from 2002 onwards and until now under the name 
`Monitoring Of Jets in Active galactic nuclei with VLBA Experiments'
(MOJAVE\footnote{\texttt{http://www.physics.purdue.edu/astro/MOJAVE/}}).
The project studies a complete sample of 135 objects 
above $-30^\circ$ observed at 15\,GHz including
dual polarisation, and its database contains images of up to
300 sources.  It consists of continuous long-term monitoring
including source-specific observing cadences, yielding high-quality
jet motions.  The well-defined sample enables 
solid statistics of the parent population 
(e.g., \cite{ars10,ars12}). 
The high-quality imaging and monitoring results have also
made possible numerous individual
source studies performed by the MOJAVE group or by others,
since all calibrated data are made publicly available.

Studies based on
the \fermi\ three-month bright source list \cite{abd09b} show that
the sources being more compact, brighter in radio, with higher
radio activity, and with
higher $\delta$ values, are favorably
detected by \fermi\ \cite{kov09b}.  The $\gamma$-ray
bright sources tend to have faster jets, especially 
in the case of quasars \cite{lis09b}.  
Concerning
the apparent opening angles, jets detected in
$\gamma$-rays tend to be broader than the non-detected
ones, but the intrinsic opening angles are similar
for detected and non-detected ones, connecting
beaming and $\gamma$-detection \cite{pus09}.
A later study shows that LAT-detected
blazars have higher $\delta$ values than the non-detected
ones, and the viewing angle distribution is different
for the $\gamma$-ray bright and weak sources;
the comoving frame viewing angle distribution
is narrower for $\gamma$-bright sources \cite{sav10}.
Furthermore, a correlation analysis showed a delay 
between the VLBI core brightness and the $\gamma$-ray
emission ($\gamma$ leads 15-GHz radio) \cite{pus10}.
Results on a joint $\gamma$-ray and radio-selected sample
show that the $\gamma$-ray loudness $G_r$ increases 
with the SED $\nu_\mathrm{IC}$, and that the
high-synchrotron-peaked (HSP\footnote{HSP show a peak in high-energy bump of 
the spectral energy distribution above $10^{15}$\,Hz, whereas intermediate- 
(ISP) and low- (LSP) synchrotron-peaked have peaks between $10^{14}$\,Hz
and $10^{15}$\,Hz, and below $10^{14}$\,Hz, respectively.
})
BL\,Lac objects have lower radio core $T_\mathrm{b}$
values \cite{lis11}.  To finish with the recently
published results, positive correlation was found between $L_\mathrm{R}$
and $L_\gamma$ \cite{ars12}. 
A study on the relationship of RM in the MOJAVE images with
$\gamma$-ray is presented in \cite{hov10} and in a future
publication.
After including optical properties, 
a positive correlation is present between $L_\mathrm{R}$ and 
the $\gamma$-ray-optical loudness for quasars,
and a negative correlation between $L_\mathrm{opt}$ and the
$\gamma$-ray-radio loudness \cite{ars12}.
A preliminary study on the relationship between the SED properties
of the MOJAVE sources and the radio properties has been also
first presented presented in \cite{cha10}, and will be published
elsewhere.

\subsection{TANAMI}
The TANAMI project (Tracking Active Galactic Nuclei with Austral Milliarcsecond 
Interferometry\footnote{\texttt{http://pulsar.sternwarte.uni-erlangen.de/tanami/}}
uses the Australian Long Baseline Array with additional telescopes to
study sources at declination below $-30^\circ$.  With a different approach,
it complements the regions of the sky not covered by MOJAVE. 
The project was started in November 2007
and observes at 8.4\,GHz and 22\,GHz.  First images at 8.4\,GHz have
been published, and a preliminary analysis shows that \fermilat-detected
sources have larger opening angles $\psi$ \cite{ojh10}.  
22\,GHz and spectral index images and  
jet kinematics will be published in a near future.

\subsection{Other studies and samples}
Some of the TeV blazars have been studied systematically with VLBI.  
One of the $\gamma$-related 
studies studies the properties of six blazars in \cite{pin10}, 
and additional ones are included in \cite{pin11}.
A VLBI-$\gamma$ study based on the Bologna Complete Sample aims to observe a 
sample of 94 nearby ($z<0.1$) sources, from which 76 are being processed
now \cite{liu11}.
A study comparing the CJF sample with the 1LAC shows a tentative correlation
between $L_\gamma$ and $\beta_\mathrm{app}$, especially for BL\,Lac objects
and $\gamma$-variable sources \cite{kar11}; the apparent speed distribution
seems to be the same for $\gamma$-detected and non-detected sources.
Another interesting study is the VLBI spectral analysis of 20 blazars
at the 1.4--15.4\,GHz range by \cite{sok10,sok11}, which data are used as 
well in connection to the $\gamma$-ray results, to test if the 
high energy emission is located at the VLBI cores.

\section{INDIVIDUAL SOURCE STUDIES}
Table~\ref{table:indiv} shows a selection of sources of special interest. 
I have tabulated sources with with published combined VLBI-$\gamma$ results,
and added some sources with more than one flare reported in ATels (to
keep the list short).
For a list of the sources detected at TeV, see 
{\scriptsize \texttt{http://tevcat.uchicago.edu/}}.
A very useful list compiled by the MOJAVE team listing all sources
being observed by the different surveys can be found at
{\scriptsize \texttt{http://www.physics.purdue.edu/astro/MOJAVE/blazarlist.html}}.

\begin{table}
\caption{Individual source studies ($\gamma$ \& VLBI) \label{table:indiv}}
\[
\centering
\resizebox{0.99\columnwidth}{!}{%
\begin{tabular}{|l|l|p{40mm}|l|l|}
\hline
\textbf{ID} 	& \textbf{Alt.\ ID} 	& \textbf{ATel}	& 	&  \\
\textbf{(B1950.0)} &			& \textbf{No.}	& \textbf{Program}$^\mathrm{a}$ 
									& \textbf{Ref}  \\
\hline
0219$+$428	& 3C\,66A		& 1753		&		& \cite{abd11d} \\ 
0235$+$164	& 			& 1744, 1784	& M12+B		& \cite{agu11a} \\
0313$+$411	& IC\,310		& 2510		& 		& \cite{kad12} \\
0316$+$413	& 3C\,84		& 2737		& M12+B		& \cite{nag10b} \\
0402$-$362	&			& 2413, 2484, 3554, 3655, 3658
							& T		& \\
0454$-$234	&			& 1898, 3703	&		& \\
0528$+$134	&			& 3412		& M12+B		& \cite{pal11} \\
0537$-$441	&			& 2124, 2454, 2591	
							& T		& \cite{hun10} \\
0716$+$714	&			& 1500, 3487, 3700	
							& M12+B		& \\
0727$-$115	&			& 1919, 2860	& M12		& \\
0805$-$077	&			& 2048, 2136	& M12		& \\
0806$+$524	&			& 1415, 3192	& M2		& \\ 
0836$+$710	& 4C\,+71.07		& 3233, 3831	& M12+B		& \\
0851$+$202	& OJ\,287		& 2256, 3680	& M12+B		& \cite{agu11b} \\
0946$+$006	& PMN\,J0948$+$0022	& 2733, 3429, 3448	
							& M2		& \cite{abd09e,fos11,gir11} \\
1101$+$384	& Mrk\,421		& ...		& M2+B		& \cite{abd11a} \\
1222$+$216	& 4C\,+21.35		& 2021, 2348, 2349, 2584, 2641, 2684, 2687	
							& M12+B		& \cite{jor11b,jor11c} \\
1226$+$023	& 3C\,273		& 1707, 2009, 2168, 2200, 2376
							& M12		& \cite{jor11c,lis11b} \\
1228$+$126	& M\,87			& 2437$^\mathrm{b}$ 		
							& M12		& \cite{acc09,abd09i,gir10b} \\
1236$+$049	&			& 1888, 3429	& M2		& \\
1253$-$055	& 3C\,279		& 1864, 2154, 2886	
							& M12+B		& \cite{jor11c} \\
1322$-$428	& Centaurus\,A		& ...		& T		& \cite{abd10b,mue11} \\
1329$-$049	& OP\,$-$050		& 2728, 2829	& M2		& \\
1343$+$451	&			& 2217, 3793	& M2		& \\
1424$-$418	&			& 2104, 2583, 3329	
							& T		& \\
1502$+$106	& OR\,103		& 1905		& M12		& \cite{abd10h} \\
1510$-$089	&			& 1743, 1897, 1968, 1976, 2033, 2385, 3470, 3473, 3694	
							& M12+B		& \cite{abd10a,mar10,ori11} \\ 
1551$+$130	& OR\,+186		& ...		& M2		& \cite{aha06,abd10y} \\
1622$-$253	&			& 2231, 2531, 3424	
							& M2		& \\
1633$+$382	& 4C\,+38.41		& 2136, 2546, 3333	
							& M12+B		& \cite{jor11a,jor11c} \\
1641$+$399	& 3C\,345		& 2316		& M12+B		& \cite{sch11,sch12} \\
1652$+$389	& Mrk\,501		& ...		& M2		& \cite{abd11c} \\ 

1803$+$784	& 			& 2386, 3322	& M12		& \\
2200$+$420	& BL\,Lac		& 2402, 3368, 3387, 3459, 3462
							& M12+B		& \cite{abd11z} \\ 
2251$+$158	& 3C\,454.3		& 1628, 1634, 2009, 2200, 2322, 2326, 2328, 2534, 2995, 3034, 3041, 3043
							& M12+B		& \cite{jor10,jor11c} \\
2345$-$167	& 			& 2408, 2972	& M12		& \\
\hline
\multicolumn{5}{l}{\footnotesize \textsc{Note:} Table updated as of February 1st, 2012} \\[-2pt]
\multicolumn{5}{l}{\footnotesize \textsc{Note:} ATels on the gravitational lens PKS\,1830$-$211 are not listed} \\[-2pt]
\multicolumn{5}{l}{\footnotesize $^\mathrm{a}$ Key: M1/2 MOJAVE 1/2; T: TANAMI; B: Boston U.} \\[-2pt]
\multicolumn{5}{l}{\footnotesize $^\mathrm{b}$ The ATel 2437 reports on post-VHE-flare eEVN observations.} \\[-2pt]
\end{tabular}
}
\]
\end{table}

Here we describe in detail a selection sources sorted by right ascension
in B1950.0 coordinates, where $\gamma$-VLBI data have been reported.

\vskip 4pt
\noindent
\textbf{AO\,0235+164}
This source has been studied in the framework of a multi-band campaign, and
multi-band light curve correlations at different bands is presented together
with VLBI analysis from the BU program \cite{agu11a}.  These results 
show hints of a new feature in the jet associated to the $\gamma$-outburst
observed by \fermi, which is interpreted as the propagation of an 
extended moving perturbation through a re-collimation structure 
at the end of the region where the jet is collimated and accelerated.

\vskip 4pt
\noindent
\textbf{IC\,310}
The galaxy 0313$+$411 in the Perseus Cluster has been
recently detected in $\gamma$-rays with an extremely hard $\gamma$-spectrum.
Sub-parsec-scale VLBA images at 8.4\,GHz detect a one-sided core-jet
structure with blazar-like radio emission oriented at the same 
position angle than the kiloparsec radio structure \cite{kad12}.  Those
findings suggest this object to have of blazar nature rather than being
a head-tail radio galaxy as it was classified in the past.

\vskip 4pt
\noindent
\textbf{3C\,84}
The \fermi-detection of 0316$+$413 in the Perseus Cluster was reported in \cite{abd09d}, including
MOJAVE data, where a brightening of the central sub-parsec-scale region is
reported, especially by comparing images from August 2008 and September 2007.
Results from 14 epochs in the 2010s carried out with the Japanese 
VLBI Network show an outburst 
associated with the central parsec near the core.  
A jet component with 
$\beta_\mathrm{app}\sim 0.23c$ is getting brighter during the $\gamma$-ray
flare, which suggests a connection between both events 
\cite{nag10b}.

\vskip 4pt
\noindent
\textbf{PKS\,0528+134}
Multi-band results during the 2nd half of 2009 show a quiescent high-energy 
behavior, and the BU program images presented a stable
state in its parsec-scale
radio jet at 43\,GHz \cite{pal11}.

\vskip 4pt
\noindent
\textbf{PKS\,0537$-$441}
First images from TANAMI are discussed by \cite{hun10}. Those results
include monitoring and spectral information between 8.4\,GHz and 22\,GHz
as compared to \fermilat\ light curves in $\gamma$-rays.

\vskip 4pt
\noindent
\textbf{OJ\,287}
Multi-epoch, multi-waveband flux and linear polarization observations of
0851$+$202 have been presented by \cite{agu11b} in the framework of
the BU blazar monitoring program.  Those observational results suggest that 
the $\gamma$-ray emission is caused by a prominent feature in the jet
$>14$\,pc away from the central engine.  The parsec-scale structure
of this source has apparently changed its direction since 2005, and the
kinematic analysis of the VLBI structure shows that two $\gamma$-flares
happen while a feature passes through a quasi-stationary shock in the jet.
Detailed results of intensive monitoring from 1995 to 2011 at 43\,GHz
are presented in \cite{agu12}, showing erratic wobbling in its sub-parsec 
jet structure.

\vskip 4pt
\noindent
\textbf{PMN\,J0948+0022}
The narrow-line Seyfert 1 source 0946$+$006 has been studied 
with a multi-band campaign tied to the
$\gamma$-flare in spring 2009, and MOJAVE and e-EVN data show a relationship
between a compact parsec-scale image with a bright core and the
$\gamma$-ray observations \cite{abd09e}.  MOJAVE results showing the
pc-scale structure and a swing in $\chi$ are presented in
the framework of the multi-band study presented in \cite{fos11}, but no
robust kinematics was possible with four epochs at the time of publication.  
The results from the eEVN yield a value of $T_\mathrm{b}=3.4\times10^{11}$\,K, 
confirming that
this object is similar to FSRQ \cite{gir11}.

\vskip 4pt
\noindent
\textbf{Mrk\,421}
A multi-band campaign on the TeV source 1101$+$384 \cite{abd11a} including VLBI data 
shows a partially resolved pc-scale radio core; the radio source showed
a low activity at all wavebands during the campaign.

\vskip 4pt
\noindent
\textbf{Centaurus\,A}
Detailed results from a multi-band campaign
on 1322$-$428
is presented in \cite{abd10b}, including TANAMI observations
where the VLBI core
size is used to calculate an upper limit on the size of the $\gamma$-ray
emitting central region ($<0.017$\,pc), whereas the slow $\beta_\mathrm{app}$
measured do not impose constrains in the value of $\Gamma$.
More detailed VLBI imaging is presented 
in \cite{mue11} where some regions 
near the core have an inverted radio spectrum, which are suggested as possible
production sites for the high energy photons, since they 
have high $T_\mathrm{b}$
and compact structure. Note that this is the only source known so far with
detectable $\gamma$-emission beyond parsec-scales.

\vskip 4pt
\noindent
\textbf{4C\,+21.35}
The BU program study on 1222$+$236 combined with multi-waveband observations
has produced interesting results on the correlation of light curves\cite{jor11b}.  
VLBI morphological results are interpreted in \cite{jor11c} as 
showing a superluminal feature crossing
at $\beta_\mathrm{app}\sim 14c$ a stationary jet fieature simultaneously with a 
$\gamma$-ray high state \cite{tan11}.

\vskip 4pt
\noindent
\textbf{M\,87}
The nearby galaxy 1228$+$126 
was detected by \fermilat\ and multi-band results
are presented in \cite{abd09i} including MOJAVE data.
The source has been intensively monitored in the mid 2000s to track the
location and nature of the high-energy and radio emission of the component HST-1
(see \cite{cha10b} and references therein).  
The relationship between the 43\,GHz VLBI brightening of the core in 2008 and
a TeV emission was presented in \cite{acc09}.
In this context, further monitoring
has been performed including the eEVN \cite{gir10b}, and the possibility that the 
$\gamma$-ray emission observed by \fermi\ comes from HST1 is still unclear.

\vskip 4pt
\noindent
\textbf{3C\,273}
A relationship between a high state in the VLBI core and a $\gamma$-ray flare
has been reported, implying a location distance of 4--11\,pc between both
emission regions \cite{lis11b}.
Observations of 1226$+$023  by the BU blazar program 
report that seven new features are present in the jet, four of
them related to $\gamma$-emission with $\beta_\mathrm{app}\sim 7-12 c$,
and a value of 0.7\,knots ejected per year is reported \cite{jor11c}.

\vskip 4pt
\noindent
\textbf{3C\,279}
The BU blazar program reports on two knots appearing in the jet of 1253$-$055, and the
time of their passage through the parsec-scale mm-wave radio core 
coincide with two prominent $\gamma$-ray events in the light curve \cite{jor11c}.
This is so far the most distant source where TeV emission has been detected.

\vskip 4pt
\noindent
\textbf{PKS\,1502+106}
A multi-band campaign including \fermilat\ and MOJAVE data has shown
the connection between $\gamma$ activity and a rotation in the EVPA
at this source \cite{abd10h}.

\vskip 4pt
\noindent
\textbf{PKS\,1510$-$089}
The source had a flare detected by \agile\ in March 2009
\cite{dam11}, also reported by \fermilat\ \cite{abd10a}.
The radio structure, spectra, and polarisation of this source are 
presented in \cite{abd10a}, 
including results from MOJAVE and from 
\cite{sok10,sok11}.
The BU blazar monitoring results
are interpreted as being caused by a bright 
knot of emission passing to the
stationary VLBI `core', which produces a long radio and X-ray outburst
lasting months after the flare \cite{mar10}.  Newer $\gamma$-flares are
interpreted as well in this picture.
An analysis of earlier MOJAVE archival data reveals that
emission at $\gamma$ and radio energies has origin in the same region \cite{ori11}.

\vskip 4pt
\noindent
\textbf{4C\,+38.41}
The source 1633$+$382 has been studied by \cite{jor11a, jor11b}.
The BU observations report that a high $\gamma$-ray state in
September 2009 is simultaneous with a high state in the VLBI core.  
The changes polarisation vectors from VLBI and the optical support the idea
of the $\gamma$-rays are connected with processes near the mm-VLBI core.

\vskip 4pt
\noindent
\textbf{3C\,345}
The source 1641$+$399 was identified as the source of $\gamma$-ray emission
reported from the 3C\,345--NRAO\,512 region \cite{sch11}.
The combination of 20-month data from \fermilat\ and 43\,GHz VLBA monitoring observations
of 3C\,345 shows that the quiescent and flaring components of $\gamma$-ray
emission are produced in a region of the jet of up to $\sim$23\,pc (deprojected),
favouring the synchrotron self-Compton mechanism for $\gamma$-ray production
\cite{sch12}.

\vskip 4pt
\noindent
\textbf{Mrk\,501} 
The source 1652$+$389 has emitted a mildly variable $\gamma$-flux
over the period 2008--2011.  A multi-band campaign of this source shows multi-frequency
VLBA data from the projects \texttt{BK150}, \texttt{BP143}, and MOJAVE; 
its VLBI core is used for the SED analysis of the source \cite{abd11c}.

\vskip 4pt
\noindent
\textbf{BL\,Lac}
The source 2200$+$420  has registered several major $\gamma$-ray flares 
detected by \fermi\, and is observed intensively in VLBI since the 1980s.
Results on a multi-band campaign are presented in \cite{abd11z}, including
VLBI images from the Fermi-related multi-wavelength campaign reported in
\cite{sok10,sok11}.  Multi-band and spectral index VLBI
images reveal a curved jet with a core region
with inhomogeneous structure and changing spectral properties rather than 
being a single, uniform, self-absorbed feature.  A turnover
by synchrotron self-absorption in the core takes place at $\sim12$\,GHz; 
and assuming $\delta=7.3$ from \cite{hov09}, 
a limit is set to the core magnetic field of $B<3$\,G.

\vskip 4pt
\noindent
\textbf{3C\,454.3}
The \fermilat\ detection of 2251$+$158 is reported in \cite{abd09c}.
The source has registered several major $\gamma$-flares 
(see e.g., the \agile\ observations in \cite{ver11}), as it seen in the
amount of ATels in table~\ref{table:indiv}.  The source has been intensively
studied with VLBI.  
The results from the BU blazar program \cite{jor10,jor11c}
report on the coincidence of $\gamma$-ray peaks with jet
features crossing the jet location at superluminal speeds.  This phenomenon
has occurred three times, coinciding with the three major flares in December 2009, April and November 2010.

%

\section{CONCLUSION}

An enormous observational VLBI effort is being performed on AGN 
to address their nature under the new light of 
$\gamma$-emission. 
Extensive studies on AGN samples and intensive
multi-band campaigns on individual sources are
underway.  As of writing this review, new 
publications are in the process of submission or revision, 
including newer correlations and observational data. 
One of the next challenges is to establish a (statistically robust) 
connection between ejection
of features in VLBI jets and $\gamma$-flares and hardness changes at
the high spectrum, once enough data have been collected during the
\fermi\ era.  The new windows opened to the Southern sky or to 
extreme regions in the parameter space such as mm-VLBI will provide
new results.  We look forward to the next $\gamma$-radio meeting
and bigger symposia for new, exciting findings.

\bigskip 
\begin{acknowledgments}
Thanks are due to K.I.\ Kellermann, M.L.\ Lister, T.\ Savolainen, and 
F.\ Schinzel for careful reading of the manuscript and very useful suggestions.
This work has made use of data from the MOJAVE database that 
is maintained by the MOJAVE team \cite{lis09a}.
The author acknowledges partial support by the Spanish MICINN through grant
AYA2009-13036-C02-02, and by the COST action MP0905 `Black Holes in a
Violent Universe'.
\end{acknowledgments}

\bigskip 

\end{document}